\documentclass[11pt,letter,oneside]{article}

\usepackage[left=.5in,right=.5in,top=1in,bottom=1in]{geometry}

\usepackage{amsmath,amssymb,amsthm}
\usepackage{graphicx}

\providecommand{\abs}[1]{\lvert#1\rvert}
\providecommand{\pdiff}[1]{\frac{\partial}{\partial #1}}
\providecommand{\pdiffarg}[2]{\frac{\partial #1}{\partial #2}}

\providecommand{\refeq}[1]{Eq.~\ref{#1}}
\providecommand{\reftwoeqs}[2]{Eqs.~\ref{#1} and \ref{#2}}

\providecommand{\reffig}[1]{Fig.~\ref{#1}}

\providecommand{\unitmatrix}[0]{\overleftrightarrow{\mathbf 1}}
\providecommand{\V}[1]{\boldsymbol #1}
\providecommand{\T}[1]{\overleftrightarrow{\boldsymbol #1}}
\providecommand{\Av}[1]{\left\langle #1 \right\rangle}
\providecommand{\C}[1]{{\cal #1}}
\providecommand{\tx}[1]{\text{#1}}

\providecommand{\kBT}[0]{k_\tx{B} T}

\renewcommand{\thesection}{\Roman{section}}
\newcommand{\erf}{\operatorname{erf}}

\newcommand{\ifcol}[2]
{
  \ifthenelse{\equal{\colopt}{col}}{#1}{#2}
}
\def\colopt{col}
\title{DNA-Protein Binding Rates: Bending Fluctuation and Hydrodynamic Coupling Effects}
\author{Yann von Hansen$^1$, Roland R. Netz$^{1 \star}$, Michael Hinczewski$^{1,2}$\\
$^1$Physics Department, Technical University of Munich, 85748 Garching, Germany\\
$^2$Institute for Physical Science and Technology, University of Maryland, College Park, MD 20742\\
$^\star$To whom correspondence should be addressed; E-mail: netz@ph.tum.de}
\date{}

\begin{document}
\maketitle
\abstract{We investigate diffusion-limited reactions between a diffusing
  particle and a target site on a semiflexible polymer, a key factor determining the kinetics of DNA-protein binding and polymerization of   cytoskeletal filaments.  Our theory focuses on two competing
  effects:  polymer shape fluctuations, which speed up association,
  and the hydrodynamic coupling between the diffusing particle and the chain, which
slows down association.  Polymer bending fluctuations are  described using a
  mean field dynamical theory, while the hydrodynamic coupling between polymer and particle is incorporated
  through a simple heuristic approximation.  Both of these we validate
  through comparison with Brownian dynamics simulations.  Neither of
  the effects has been fully considered before in the biophysical
  context, and we show they are necessary to form accurate estimates
  of reaction processes.  The association rate depends on the
  stiffness of the polymer and the particle size, exhibiting a maximum for intermediate
  persistence length and a minimum for intermediate particle radius. In the parameter range relevant to DNA-protein
  binding, the rate increase is up to $100\%$ compared to the
  Smoluchowski result for simple center-of-mass motion.  The
  quantitative predictions made by the theory can be tested experimentally.}
\clearpage

\section{Introduction}
\label{Sec:Intro}

Reactions between semiflexible polymers and small molecules are
ubiquitous in cells, playing a crucial role in a large number of
biological processes: examples include the interaction of
gene-regulating proteins with specific target sites on DNA, and the
polymerization of DNA or structural proteins such as actin and
tubulin.  Many of these reactions are diffusion-limited~\cite{Berg1985}---the
activation free-energy is negligible compared to the thermal energy
$\kBT$ and the reaction is not hindered by other steric or conformational factors---so the overall association speed therefore depends on the
rate at which the reactive molecules approach each other.

DNA-protein interaction has been the most widely studied process of
this type~\cite{Halford2004,Bruinsma2002}, attracting attention since the first
measurement of the reaction rate between the \textit{lac} repressor and
operator~\cite{Riggs1970} revealed that it far exceeds the 3D
diffusion-limit. The quest to identify the underlying mechanism
culminated in the seminal idea of facilitated diffusion by Berg,
Winter, and von Hippel~\cite{Berg1981,Winter1981a,Winter1981b}, but
the general description of DNA-protein interaction is still far from
complete.  Although recent single molecule experiments show evidence
that certain proteins indeed make use of facilitated diffusion
\cite{Wang2006,Bonnet2008}, the majority of measured reaction rates
for DNA-binding proteins~\cite{Halford2004,Kleinschmidt1988,Halford2009} do not exceed the Smoluchowski result
for 3D diffusion~\cite{Smoluchowski1917}.  Thus recent
years have seen extensive theoretical efforts~\cite{Slutsky2004,Lomholt2005,Cherstvy2008,Florescu2009,Lomholt2009} revisiting the underlying assumptions of facilitated diffusion, and examining it anew in response to experimental advances.

Diffusion-controlled reactions with targets on flexible polymers are a
well-established subject in polymer physics, with general analytical
frameworks developed by Wilemski and Fixman
~\cite{Wilemski1974a,Wilemski1974b}, and Szabo, Schulten, and
Schulten~\cite{Szabo1980} decades ago.  Using the Wilemski-Fixman
approach and a Gaussian model for circular polymers, a study by Berg looked at the influence of internal DNA
motion on the binding of proteins~\cite{Berg1984}.  But little is
known about the role of chain stiffness, and both Berg's study
and the classic polymer reaction rate theories do not include
hydrodynamics.  This is a significant oversight, because
solvent-mediated interactions modify not only the approach of the
particle to the target site, but also the fluctuation of the entire
polymer contour, and thus all time scales involved in the
dynamics.  Moreover, the two effects are in opposition: polymer
fluctuations lead to the target site exploring a larger
configurational space than simple center-of-mass diffusion, and hence
the association rate is enhanced; the hydrodynamic coupling between
the particle and coil, on the other hand, reduces their relative
mobility, decreasing the rate.  The observed binding rate is a subtle
competition between these two phenomena.  As experiments are providing
an ever more detailed picture of biological reactions at the single
molecule level, we need theories that begin to grapple with the full
complexity of polymer-particle diffusive motion.

The paper is organized as follows:
in Sec.~\ref{Sec:Theory} we briefly review the renewal approach used to derive first-passage times and association rates in the diffusion-controlled limit; a discussion of a few effects going beyond this limit is contained in \ref{Sec:AddEffects}. The theoretical analysis of the dynamics in the polymer-particle system makes use of a mean field approach for the polymer dynamics and a heuristic approximation for inter-particle hydrodynamics. The main results are shown in Sec.~\ref{Sec:DynamicGF}, while a brief review of the theory and calculational details are postponed to the appendices~B and C.
The Brownian dynamics (BD) simulation method used to independently test the theoretical description is described in Sec.~\ref{Sec:BD}.
The detailed validation of the theory through BD simulation results is presented in Sec.~\ref{Sec:Results}.  We then investigate how the mechanical characteristics of the semiflexible polymer---the contour and persistence lengths---affect the reaction rate.  Ultimately our theory yields quantitative predictions that can be tested empirically.
Sec.~\ref{Sec:Discussion} summarizes the main results and offers suggestions for future experiments.

\section{Theory}
\label{Sec:Theory}
\subsection*{Diffusion-controlled reaction rates}
\label{Sec:FPTandRates}
In general, association rates are calculated by finding the steady-state solution of a diffusion equation with absorbing boundary conditions~\cite{Collins1949, Wilemski1973}. In our case, we can employ the simpler renewal approach~\cite{Cox, vanKampen}, which avoids the necessity of imposing absorbing boundaries: first passage times and binding rates can be directly extracted from the solution of the unbounded problem, i.e. the dynamic Green's function describing the time evolution of the relative distance between reactants subject to diffusive motion.
The renewal approach is in principle equivalent to the Wilemski-Fixman formalism, and they share the same underlying approximations~\cite{Likthman2006,Eun2007}: (i) the Green's function for the time evolution of the particle-target distance is assumed to be that of a stationary Markov process; (ii) excluded-volume between the reactants is ignored.  In the following we briefly review the main aspects of the approach.

For a stochastic process in one dimension, the Green's function $g(x,x_0;t)$ specifies the probability to find the particle at $x$ at time $t$ given a starting point $x_0$ at time $t=0$. If we are interested in paths that reach a boundary point $x_\tx{a}$ in time $t$, the corresponding transition probability can be written as a convolution,
\begin{equation}
\label{Eq:Renewal}
g(x_\tx{a},x_0;t)=\int_0^t
dt^\prime\,f(t^\prime;x_\tx{a},x_0)\,g(x_\tx{a},x_\tx{a};t-t^\prime),
\end{equation}
where $f(t;x_\tx{a},x_0)$ is the first passage time distribution: the probability of reaching $x_\tx{a}$ in time $t$ starting from $x_0$ at $t=0$ without passing through $x_\tx{a}$ along the way.  A Laplace transformation $\C L$ acting on the time variable $t$ leads to
\begin{equation}
\label{Eq:RenewalLT}
\tilde g(x_\tx{a},x_0;s)=\tilde f(s;x_\tx{a},x_0)\tilde g(x_\tx{a},x_\tx{a};s)\quad\Leftrightarrow\quad f(t;x_\tx{a},x_0)={\cal L}^{-1}\left[\frac{\tilde g(x_\tx{a},x_0;s)}{\tilde g(x_\tx{a},x_\tx{a};s)}\right],
\end{equation}
where Laplace transforms are denoted by $\C L[f(t)]=\tilde f(s)$.
Note that $g(x,x_0;t)$ is the Green's function in the absence of
absorbing boundary conditions; we assume only that the probability
vanishes at $x \rightarrow \pm \infty$.

To investigate the role of polymer fluctuations and of hydrodynamics on the reaction rates of diffusion controlled reactions,  we consider an idealized situation: the only parameter relevant for the binding process then is the radial distance $r$ between the reactants meaning that the problem is effectively reduced to one dimension.
Deviations from this simple picture and their impact on the association rates are discussed in \ref{Sec:AddEffects}. 
In the case of perfect absorption, particles bind when they collide for the first time---when the relative distance $r$ reaches the absorption radius $r_\tx{a}$.
As a result, the binding rate $k_\tx{a}(t)$ can be obtained from the first passage time distribution by integrating over all possible initial separations:
\begin{equation}
\label{Eq:RenewalRates}
k_\tx{a}(t)=4\pi\int_{r_\tx{a}}^\infty dr_0\,r_0^2\,f(t;r_\tx{a},r_0).
\end{equation}
The steady-state rate $k_\tx{a}$ reached at long times is extracted from
the Laplace transform $\tilde k_\tx{a}(s)$ using the final value theorem:
\begin{equation}
\label{Eq:RenewalAsymptoticRates}
k_\tx{a}=\lim_{t\rightarrow\infty} k_\tx{a}(t)=\lim_{s\rightarrow 0} s\, \tilde k_\tx{a}(s).
\end{equation}
For the simple case of two kinds of uncoupled Brownian particles with
diffusion constants $D_1$ and $D_2$, the radial Green's function,
which will explicitely be shown in the next section (Eqs.~\ref{Eq:GFrad}-\ref{Eq:VarGFn2Par}), 
can be easily written down and the result for the first passage time distribution in Laplace space reads
\begin{equation}
\label{Eq:LaplaceTransFormFirstPassageTimeBrownianMotion}
\tilde f(s;r_\tx{a},r_0)=\frac{r_\tx{a}}{r_0}\exp{\left(-(r_0-r_\tx{a})\sqrt{\frac{s}{D}}\right)},
\end{equation}
with total diffusion constant $D=D_1+D_2$.
Applying \refeq{Eq:RenewalRates}, which corresponds to imposing uniform initial concentrations at time $t=0$, one obtains the association rate:
\begin{equation}
\label{Eq:RateDS}
k_\tx{a}(t)=4 \pi  D r_\tx{a}\left(1+\frac{r_\tx{a}}{\sqrt{\pi D t}}\right),
\end{equation}
which at long times reduces to the well-known
Smoluchowski rate~\cite{Smoluchowski1917} $k_\tx{S}=4\pi D r_\tx{a}$ constituting the upper limit for reaction rates governed by Brownian diffusion in three dimensions.

For the more complicated polymer-particle case, the renewal method works analogously, but the Laplace transforms must be carried out numerically using standard numerical techniques~\cite{GSL}: the Laplace-transformation of the numerator in \refeq{Eq:RenewalLT} is performed with the routine \textit{gsl\_integration\_qags} with upper integration boundary $50/s$, while the Laplace transform of the denominator in \refeq{Eq:RenewalLT} and the numerical integration over initial separations $r_0$ in \refeq{Eq:RenewalRates} are obtained using the routine \textit{gsl\_integration\_qagiu}; a workspace of $10^5$ intervals was allocated for all these routines and the relative and absolute error bounds were set to $10^{-7}$. For the numerical implementation of the final value theorem in \refeq{Eq:RenewalAsymptoticRates} a value $s\approx 1-2\cdot10^{-8}\tau^{-1}$ ($\tau\equiv6\pi\eta a^3/\kBT$ being the diffusional time scale of a spherical particle of radius $a$ in a solvent of viscosity $\eta$) proved appropriate: within this range the numerical evaluation of the steady state Smoluchowski-rate $k_\tx{S}$ coincided with the analytic result within a typical relative accuracy of $\lesssim 10^{-5}$. Results of the renewal approach applied to DNA-protein dynamics are presented in Sec.~\ref{Sec:Rates}.

\subsection*{Dynamic Green's functions}
\label{Sec:DynamicGF}
As outlined above the knowledge of dynamic Green's functions specifying the probability that a radial distance $r$ between two objects is realized at time $t$ after starting at a radial distance $r_0$ allows the calculation of association rates.

The first step towards the description of the polymer-particle system consists in the analysis of the polymer motion itself: 
The dynamics of semiflexible polymers including hydrodynamics can be captured by a mean field theory discussed in detail in Ref.~\cite{Hinczewski2009}, where the model was shown to provide an accurate description of internal polymer kinetics, validated through extensive comparisons with BD simulations. Moreover, the theory can be independently tested by comparison to recent fluorescence correlation spectroscopy experiments~\cite{Petrov2006}: without any fitting parameters, it exhibits excellent agreement for the mean square displacement (MSD) of tagged ends of single dsDNA fragments diffusing in solution~\cite{Hinczewski2009c}.

Here we demonstrate that the mean field approach in addition also exhibits excellent agreement for the dynamic Green's function characterizing the motion of specific points on the polymer contour when comparing theory and BD results. A short overview of the theory and details concerning the derivation can be found in \ref{Sec:MFT}; the result of the calculation involving a normal mode expansion is a Gaussian Green's function specifying the conditional probability that a point on the polymer contour (specified by the arc-length variable $s$) reaches spatial position $\V r$ in time $t$ given a start at $\V r_0$:
\begin{align}
\label{Eq:GF3D}
G(\V r,\V r_0;t)=&\left(2\pi V(t)\right)^{-3/2}\,\exp\left(-\frac{(\V r -\V r_0)^2}{2V(t)}\right),\\
\label{Eq:VarGFPol}
V(t)=&2 D_\tx{pol} t + 2 \kBT \sum_{n=1}^{N}\frac{\Theta_n}{\Lambda_n}\left(1-e^{-\Lambda_n t}\right)\Psi_n(s)^2.
\end{align}
In Eq.~\ref{Eq:VarGFPol} $D_\tx{pol}$ denotes the center-of-mass diffusion constant of the polymer coil and is given in \ref{Sec:MFT}. The values of the fluctuation dissipation parameters $\Theta_n$ and of the inverse relaxation times $\Lambda_n$ result from the (numerical) evaluation of the hydrodynamic interaction tensor; the normal modes $\Psi_n(s)$ with mode number $n$ additionally depend on polymer parameters, i.e. contour length $L$ and persistence length $l_\tx{p}$.
The variance $V(t)$ in Eq.~\ref{Eq:VarGFPol} thus has contributions of the center-of-mass motion of the entire polymer coil and of internal fluctuations of the contour: on large time scales $t\gg\Lambda_1^{-1}$ the variance reduces to $V(t) \approx 2 D_\tx{pol} t$ and hence the motion is dominated by the Brownian diffusion of the polymer's center-of-mass, for smaller times, however, the contribution from internal polymer modes becomes important. The radial Green's function $G_\tx{rad}(r;t)$ for a particle starting at $\V r_0=0$ is obtained by integrating $G(\V r,0;t)$ over the surface of a sphere of radius $r$:
\begin{equation}
\label{Eq:GFradPol}
G_\tx{rad}(r;t)=\frac{4\pi r^2}{\left(2\pi V(t)\right)^{3/2}}\,\exp{\left(-\frac{r^2}{2V(t)}\right)}.
\end{equation}
In Sec.~\ref{Sec:Pol} we compare the time evolution of this transition probability for the case of the end-monomer ($s=\pm L/2$) to BD simulation results.

Hydrodynamic interactions influence the relative motion of diffusing objects: these interactions are included in our analytic approach by a heuristic approximation, which is motivated for the case of two spherical particles and for the case of a single diffusing particle and a specific target site on a polymer in \ref{Sec:Hydro}.

In general, the dynamic Green's function takes the form of a Gaussian centered at the initial separation $\V r_0$; the integration over a sphere of radius $r=\abs{\V r}$ leads to the radial Green's function:
\begin{equation}
\label{Eq:GFrad}
G_\tx{rad}^\alpha(r,r_0;t)=\frac{r}{r_0 \sqrt{2\pi V^\alpha(t)}}
\left[\exp{\left(-\frac{(r-r_0)^2}{2V^\alpha(t)}\right)}-\exp{\left(-\frac{(r+r_0)^2}{2V^\alpha(t)}\right)}\right],
\end{equation}
where the superscript $\alpha=\tx{n},\tx{h}$ discriminates between radial Green's functions without and with hydrodynamics. In total, we distinguish four cases in our analysis: two spherical particles without hydrodynamics (i) and with hydrodynamics (ii), and a spherical particle and a specific target point on a semiflexible polymer without hydrodynamics (iii) and with hydrodynamics (iv), for which the respective variances are:
\begin{align}
\label{Eq:VarGFn2Par}
\tx{(i)}&\qquad V^\tx{n}(t)=2(D_1+D_2)t,\\
\label{Eq:VarGFh2Par}
\tx{(ii)}&\qquad V^\tx{h}(t,r_0)=2(D_1+D_2-2\bar\chi(r_0,t))t,\\
\label{Eq:VarGFnPolPar}
\tx{(iii)}&\qquad V^\tx{n}(t)=2 D_\tx{par}t+2 D_\tx{pol} t + 2 \kBT \sum_{n=1}^{N}\frac{\Theta_n}{\Lambda_n}(1-e^{-\Lambda_n t})\Psi_n(s)^2,\\
\label{Eq:VarGFhPolPar}
\tx{(iv)}&\qquad V^\tx{h}(t,r_0)=2 D_\tx{par}t+2 D_\tx{pol} t + 2 \kBT \sum_{n=1}^{N}\frac{\Theta_n}{\Lambda_n}(1-e^{-\Lambda_n t})\Psi_n(s)^2-4\bar\chi(r_0,t)t,
\end{align}
where $D_1$ and $D_2$ are the diffusion constants in the case of relative diffusion of two spherical particles, $D_\tx{par}$ is the diffusion constant of the free particle in the case of the polymer-particle reaction, $D_\tx{pol}$ is the polymer's center-of-mass diffusion constant and the sums over the mode numbers $n$ again characterize the polymer's contour fluctuations (compare Eq.~\ref{Eq:VarGFPol}). The slowing down of the relative motion due to hydrodynamics is reflected by the effective coupling parameter $\bar\chi(r_0,t)$, which is motivated and shown in its full functional form in \ref{Sec:Hydro}.
The non-hydrodynamic and hydrodynamic Green's functions for the two systems we consider are compared to BD results in Sec.~\ref{Sec:Results}.

\section{Brownian dynamics simulations}
\label{Sec:BD}
To test the analytic results of Sec.~\ref{Sec:Theory}, we simulate a semiflexible polymer in solution adopting a standard Brownian
dynamics scheme~\cite{Ermak1978}, in which the polymer is modeled as a
chain of $M$ beads of radius $a$. For the low Reynolds number regime,
the Langevin equation governing the time evolution of the position $\V
r_i$ of bead $i$ is given by
\begin{equation}
\label{Eq:LangevinBD}
\frac{d\V r_i(t)}{dt}=\sum_{j=1}^M\T\mu_{ij}\cdot
\underbrace{\left(-\frac{\partial U(\V{r}_1,\dots,\V{r}_M)}{\partial
    \V{r}_j}\right)}_{\V f_j(t)}+\V{\xi}_i(t).
\end{equation}
The long-range hydrodynamic interactions---the fact that a force $\V
f_j$ acting on bead $j$ creates a flow-field affecting the motion of
bead $i$---are described by the Rotne-Prager mobility matrix
$\T\mu_{ij}$~\cite{Rotne1969}:
\begin{align}
\label{Eq:HMatrixBD}
\T{\mu}_{ij}&=\mu_0\delta_{i j}\unitmatrix+(1-\delta_{i j})\T\mu(\V r_{ij}),\\
\label{Eq:RotnePragerTensor}
\T\mu(\V r_{ij})&=\frac{1}{8\pi\eta
  r_{ij}}\left[\unitmatrix+\frac{\V{r}_{ij}\otimes\V{r}_{ij}}{r_{ij}^2}\right]+\frac{a^2}{4\pi\eta
  r_{ij}^3}\left[\frac{\unitmatrix}{3}-\frac{\V{r}_{ij}\otimes\V{r}_{ij}}{r_{ij}^2}\right],
\end{align}
where $\V r_{ij} = \V r_i - \V r_j$, $r_{ij}=\abs{\V r_{ij}}$, $\unitmatrix$ is the $3\times 3$
identity matrix, and $\mu_0=(6\pi\eta a)^{-1}$ is the Stokes
self-mobility of a sphere of radius $a$ in a solvent of viscosity
$\eta$. The stochastic contributions $\V{\xi}_i(t)$ in Eq.~\ref{Eq:LangevinBD} are assumed to be Gaussian random vectors which are
hydrodynamically correlated according to the fluctuation-dissipation
theorem:
\begin{equation}
\label{Eq:CorrelationBD}
\Av{\V{\xi}_i(t)\otimes\V{\xi}_j(t')}=2\kBT\,\T\mu_{ij}\,\delta(t-t').
\end{equation}
The inter-bead potential $U=U_{WLC}+U_{LJ}$ determining the
configuration-dependent forces, where
\begin{equation}
\label{Eq:PotentialsBD}
\begin{split}
U_\tx{WLC}&=\frac{\gamma}{4a}\sum_{i=1}^{M-1}(r_{i+1,i}-2a)^2+\frac{\kappa}{2a}\sum_{i=2}^{M-1}(1-\cos
\theta_i),\\ 
U_\tx{LJ}&=w\sum_{i<j}\Theta(2a-r_{ij})\left[\left(\frac{2a}{r_{ij}}\right)^{12}-2\left(\frac{2a}{r_{ij}}\right)^6+1\right],
\end{split}
\end{equation}
consists of a shifted harmonic potential between adjacent beads of
strength $\gamma = 200\kBT /a$, a bending potential of strength
$\kappa = l_\tx{p} \kBT$ between adjacent bonds, and a pairwise truncated
Lennard-Jones potential $U_\tx{LJ}$ of strength $w = 3\kBT$. Here
$\theta_i$ is the angle between the bond vectors $\V r_{i,i-1}$ and
$\V r_{i+1,i}$.  The first term in the worm-like chain potential
$U_\tx{WLC}$ keeps the contour length $L$ approximately fixed, while the second one with modulus $\kappa$ takes care of the bending stiffness of the chain.  The
repulsive Lennard-Jones potential $U_\tx{LJ}$ prevents significant
bead overlap, which is a source of numerical instabilities.  In
Sec.~\ref{Sec:PolPar}, where we model the motion of a free particle
(i.e. a protein or free monomer) relative to the polymer chain, the
particle is represented by an additional bead, not connected to the polymer chain, and subject only to hydrodynamic interactions between the
particle and the chain. We do not account for steric interactions between particle and polymer in order to simplify the comparison with the theoretical results where such effects cannot be properly included. A short discussion of excluded-volume effects is given in \ref{Sec:AddEffects}. To avoid numerical instabilities in situations
where the free particle overlaps with the polymer, in this case we use
the Rotne-Prager-Yamakawa tensor~\cite{Yamakawa1970}, which modifies
\refeq{Eq:RotnePragerTensor} for overlapping beads:
\begin{equation}
\T\mu(\V r_{ij})=\mu_0\left[\left(1-\frac{9r_{ij}}{32a}\right)\unitmatrix+\frac{3r_{ij}}{32a}\, \frac{\V r_{ij}\otimes\V r_{ij}}{r_{ij}^2}\right]\quad \text{if}\:\: r_{ij}\leq 2a.
\end{equation}

\refeq{Eq:LangevinBD} is discretized and integrated numerically using
the Euler algorithm. The correlated stochastic contributions of
\refeq{Eq:CorrelationBD} are obtained from uncorrelated Gaussian noise
by means of a Cholesky decomposition of the hydrodynamic matrix
$\T\mu_{ij}$. In all the results below, lengths are measured in units
of $a$, energies in units of $\kBT$ and time in units of
$\tau=a^2/(\kBT\mu_0)$. The time step is $\Delta t=3\times10^{-4}\:
\tau$, and a typical simulation lasts $10^9$ steps, after an initial
thermalization period of $10^6-10^7$ steps.  To reduce computational
costs the Cholesky decomposition is only performed every 5 time steps.
For a given chain length $L=2a(M-1)$ and persistence length $l_\tx{p}$, the
quantities of interest are averaged over $25-2500$ different
trajectories until the convergence is satisfactory.

\section{Results}
\label{Sec:Results}

In order to validate the various analytical approaches described in
Sec.~\ref{Sec:Theory}, we test them against BD simulations.  Since the
binding rates depend sensitively on having good estimates for the
Green's functions, we will focus on showing that the MFT approximation
for the transition probabilities can reproduce the crucial physical
effects: (i) the influence of internal polymer modes on the diffusion
of the target site; (ii) the slow-down of relative motion between the
free particle and target due to hydrodynamics.  For simplicity, we
concentrate in our analysis on one particular target site, the
end-monomer of the chain.  This has special relevance in biological
processes, for example in the case of polymerization. Nevertheless, our theoretical approach is equally applicable to other target positions along the polymer contour: the results are similar though changes in the association rates are clearly less drastic because of the reduced mobility of target sites in the middle of the chain compared to the one at the chain's end.

\subsection*{Polymer motion}
\label{Sec:Pol}

\begin{figure}
   \begin{center}
      \includegraphics*[width=3.375in]{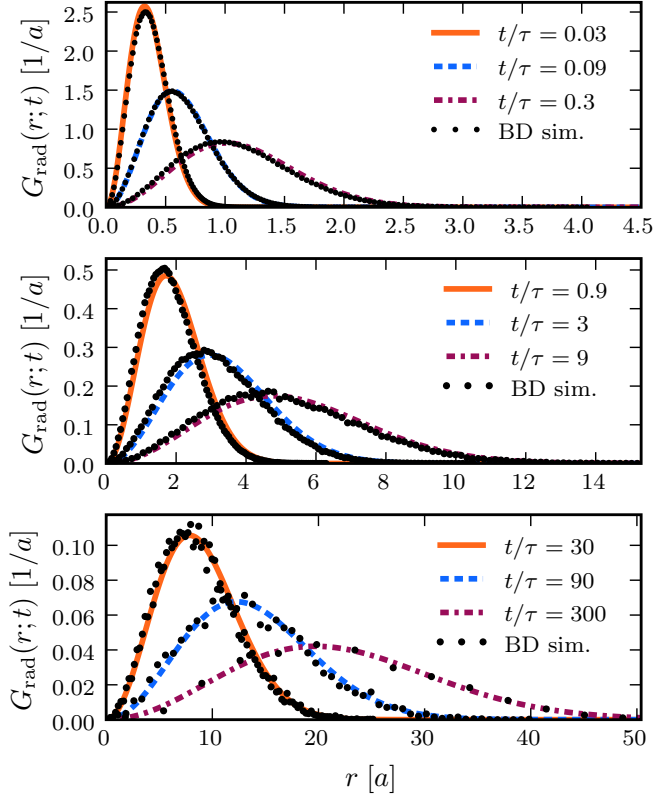}
      \caption{The probability $G_\tx{rad}(r,t)$ that the end-point
        of a polymer with length $L=100a$ and persistence length
        $l_\tx{p}=20a$ will diffuse a distance $r$ in time $t$.  The MFT
        predictions (Eqs.~\ref{Eq:VarGFPol} and \ref{Eq:GFradPol}, \textit{lines}) are compared to the results from hydrodynamic BD simulations (\textit{black dots}) for different times $t$ measured in units of $\tau=a^2/(\kBT\mu_0)$.}
      \label{Fig:Polymer_s50_N14}
   \end{center}
\end{figure}

\begin{figure}
   \begin{center}
      \includegraphics*[width=3.375in]{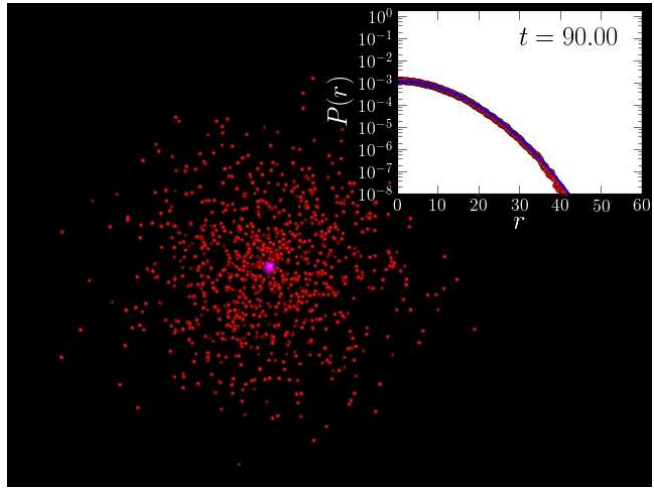}
      \caption{The movie shows the motion of polymer end-points (\textit{red dots}) superimposed from an ensemble of simulation trajectories, each starting at the origin (\textit{purple dot}). The inset shows the evolving radial probability
$P(r,t)=G_\tx{rad}(r;t)/r^2$ taken from the simulation histograms (\textit{red markers}) compared to the analytic MFT results of Eqs.~\ref{Eq:VarGFPol}-\ref{Eq:GFradPol} (\textit{blue curve}).}
      \label{Fig:Movie}
   \end{center}
\end{figure}

We begin by considering just the internal relaxation of a polymer of total contour length $L$:
the radial Green's function of Eqs.~\ref{Eq:VarGFPol}-\ref{Eq:GFradPol} describing the diffusive motion of the polymer end-point ($s=\pm L/2$).  In \reffig{Fig:Polymer_s50_N14} we compare this analytical MFT expression for $G_\tx{rad}(r,t)$ to histograms extracted from simulations of a chain with $L=100a$, $l_\tx{p} = 20a$.
The histograms are based on the
analysis of 25 independent trajectories, each with $10^8$ steps.  The
time evolution of the probability distribution is in excellent
agreement with $G_\tx{rad}(r,t)$ (Eqs.~\ref{Eq:VarGFPol} and \ref{Eq:GFradPol}) over time scales spanning four
orders of magnitude.  Note that the largest time-scale considered ($t
= 300\,\tau$) is less than the largest relaxation time of the polymer,
$\Lambda_1^{-1}\sim 2\times 10^3\,\tau$, meaning that internal fluctuations
dominate the polymer motion throughout this entire time range; the time evolution is illustrated more directly in the movie linked in Fig.~\ref{Fig:Movie} showing the remarkable agreement, even in the tails of the distribution (which
are visible on the logarithmic scale spanning seven decades of magnitude).
Our comparison here is more detailed than in the earlier
MFT study of Ref.~\cite{Hinczewski2009}, since we consider the full
transition probability and not just the end-point MSD.  But the
conclusion is the same: the MFT provides a highly accurate picture of
internal polymer dynamics.

\subsection*{Relative motion of two spherical particles including hydrodynamics}
\label{Sec:2Par}

\begin{figure}
   \begin{center}
      \includegraphics*[width=3.375in]{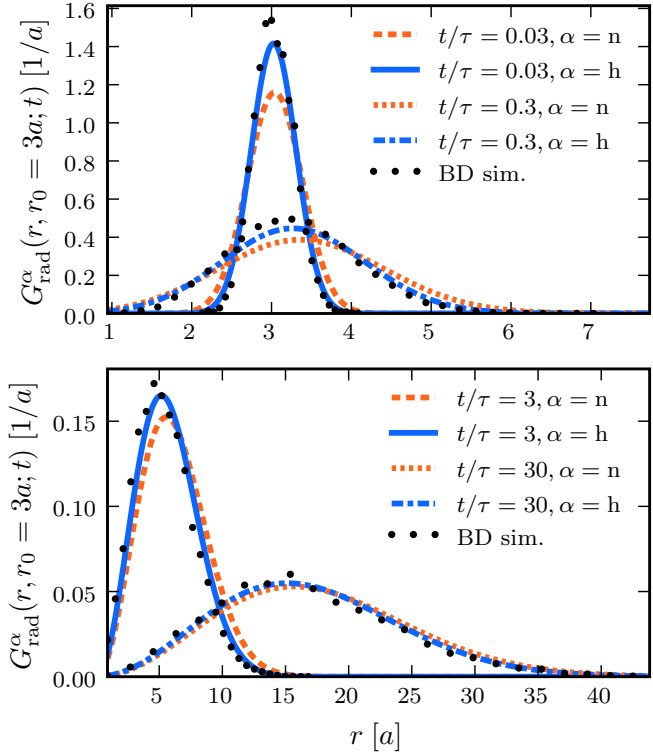}
      \caption{	The probability $G_\tx{rad}^{\alpha}(r,r_0;t)$ that two
        spherical particles of radius $a$ end up at separation $r$
        after time $t$ given an initial separation $r_0 = 3a$.
	For each time $t$, BD simulation data including hydrodynamics
        (\textit{black dots}) are compared to two theoretical results: the
        non-hydrodynamic Green's function for
        decoupled particles labeled by $\alpha=\text{n}$ (\reftwoeqs{Eq:GFrad}{Eq:VarGFn2Par}), and the approximate
        hydrodynamic Green's function labeled by $\alpha=\text{h}$ (\reftwoeqs{Eq:GFrad}{Eq:VarGFh2Par}).  Times are measured in units of $\tau=a^2/(\kBT\mu_0)$.}
      \label{Fig:2Par_r0_3}
   \end{center}
\end{figure}

\begin{figure}
   \begin{center}
      \includegraphics*[width=3.375in]{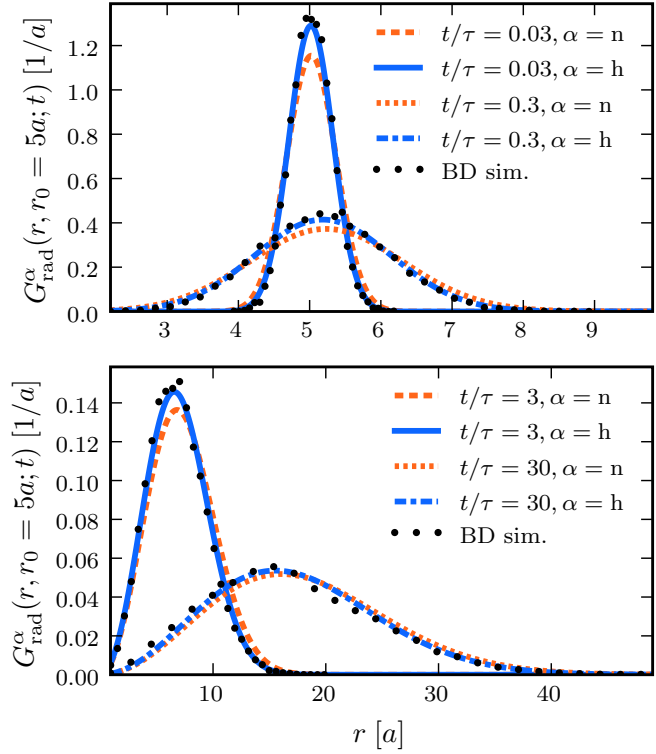}
      \caption{      Same as in \reffig{Fig:2Par_r0_3}, but with initial
        separation $r_0=5a$ between the beads.}
      \label{Fig:2Par_r0_5}
   \end{center}
\end{figure}

Since the heuristic estimate for the hydrodynamic Green's function
$G_\tx{rad}^\tx{h}(r,r_0;t)$ of two freely diffusing particles
(Eqs.~\ref{Eq:GFrad} and \ref{Eq:VarGFh2Par}) is the basis of our
approach to polymer-particle interactions, we need to check that this
estimate is reasonable.  For this purpose, we performed hydrodynamic BD simulations
of two single spheres of radius $a$ and extracted histograms from the
variation of the inter-bead distance over time.  In total 2400 independent
trajectories were calculated, each starting at a distance $r_\tx{init}=3a$ and
lasting $10^7$ time steps. For given separations $r_0$ and $r$,
relevant transition events were identified, i.e.~parts of the
trajectories starting at a certain distance $r_0\pm\delta r/2$ and
ending at $r \pm \delta r/2$, where $\delta r$ is the histogram
binwidth.  These were used to estimate the probability distribution
corresponding to $G_\tx{rad}^\tx{h}(r,r_0;t)$.  In
\reffig{Fig:2Par_r0_3} we show the BD simulation
results for $r_0=3a$ and various $t$ together with
$G_\tx{rad}^\tx{h}$ (Eqs.~\ref{Eq:GFrad} and \ref{Eq:VarGFh2Par}) and the hydrodynamically decoupled Green's function
$G_\tx{rad}^\tx{n}$ (Eqs.~\ref{Eq:GFrad} and \ref{Eq:VarGFn2Par}). Analogous results for an initial separation $r_0=5a$ are shown in \reffig{Fig:2Par_r0_5}.
The inhibition of relative motion due to hydrodynamics is clearly evident at short times: the distributions from the simulations are more
narrowly peaked than the decoupled results, which spread out
noticeably faster.  The slowing down is correctly reproduced in the
$G_\tx{rad}^\tx{h}$ curves, though it is slightly underestimated
when compared to the simulation data.  Despite the simplicity of the
underlying approximation, $G_\tx{rad}^\tx{h}$ still captures the
essential features of the interaction.  For longer times, when the
inter-particle distance has become large and hydrodynamics plays a
smaller role, $G_\tx{rad}^\tx{h}$ converges to
$G_\tx{rad}^\tx{n}$, and both agree with the BD results.

\subsection*{Relative motion of a polymer and a spherical particle including hydrodynamics}
\label{Sec:PolPar}

\begin{figure}
   \begin{center}
      \includegraphics*[width=3.375in]{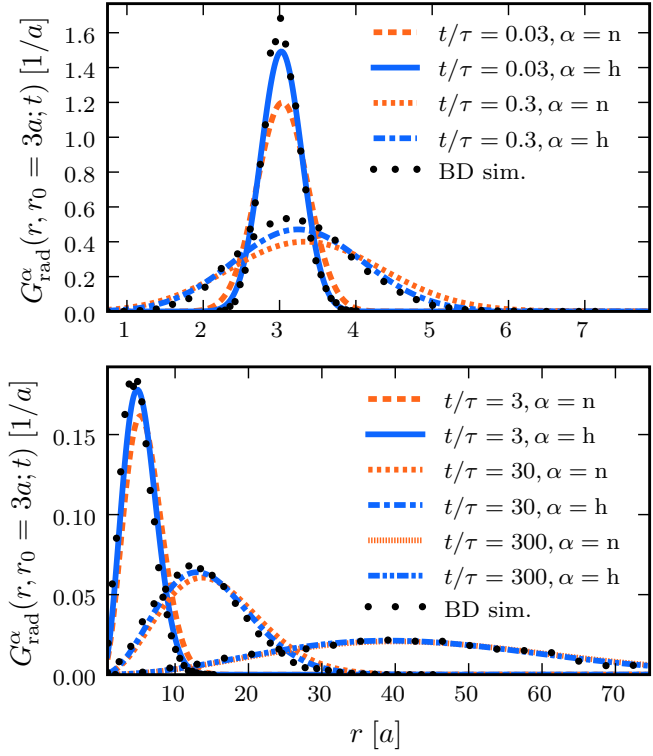}
      \caption{	The probability $G_\tx{rad}^{\alpha}(r,r_0;t)$ that the
        end-monomer of a polymer with $L=100a$, $l_\tx{p} = 20a$, and a
        spherical particle of radius $a$ end up at separation $r$
        after time $t$ given an initial separation $r_0 = 3a$. For each time $t$, BD simulation data including hydrodynamics (\textit{black dots}) are compared to
        two theoretical results: the non-hydrodynamic Green's function labeled by $\alpha=\text{n}$ (\reftwoeqs{Eq:GFrad}{Eq:VarGFnPolPar}), and the
        approximate hydrodynamic Green's function labeled by $\alpha=\text{h}$ (\reftwoeqs{Eq:GFrad}{Eq:VarGFhPolPar}).  Times are measured in units of $\tau=a^2/(\kBT\mu_0)$.}
      \label{Fig:PolPar_r0_3}
   \end{center}
\end{figure}

\begin{figure}
   \begin{center}
      \includegraphics*[width=3.375in]{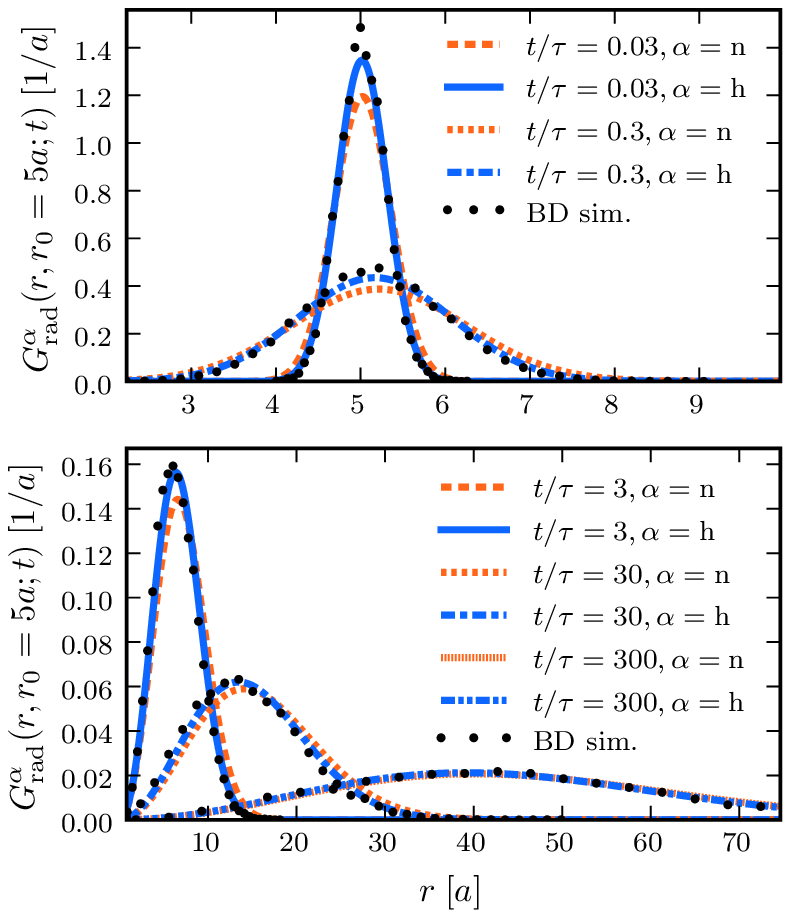}
      \caption{Same as in \reffig{Fig:PolPar_r0_3}, but with initial separation $r_0=5a$ between the end-monomer and the free particle.}
      \label{Fig:PolPar_r0_5}
   \end{center}
\end{figure}

The final test for our Green's function approach concerns the
approximate description of polymer-particle relative motion, contained
in Eqs.~\ref{Eq:GFrad} and \ref{Eq:VarGFhPolPar}.  To extract the
corresponding transition probabilities from BD simulations, we take a
polymer with $L=100a$, $l_\tx{p}=20a$, allow it to thermalize, and then add
a single free particle of radius $a$ that has no excluded-volume, but is hydrodynamically coupled to the polymer beads. The particle is positioned at
an initial distance $r_\tx{init} = 3a$ from one of the polymer ends, and data
is collected over $10^7$ time steps.  This procedure is repeated to
obtain 2500 independent trajectories.  As before,
relevant transition events are used to determine the probability
distribution for diffusing from a separation $r_0$ to $r$ in time $t$.
The results, plotted in \reffig{Fig:PolPar_r0_3} for $r_0=3a$ and in \reffig{Fig:PolPar_r0_5} for $r_0=5a$, are qualitatively similar to the case of two particles: compared to the non-hydrodynamic case we see a reduced relative mobility between the target site and the particle
for small separations.
 The hydrodynamic Green's function is quite close to the simulation results, again slightly underestimating the strength of the coupling.  However, given the complexity of the correlations between the particle and the entire polymer coil, it is notable that we are able to get good quantitative agreement.  Since the association rates are derived directly from this Green's function, we conclude that we should be able to obtain realistic estimates for the reaction process.

\subsection*{Association rates}
\label{Sec:Rates}

\begin{figure}
   \begin{center}
      \includegraphics*[width=3.375in]{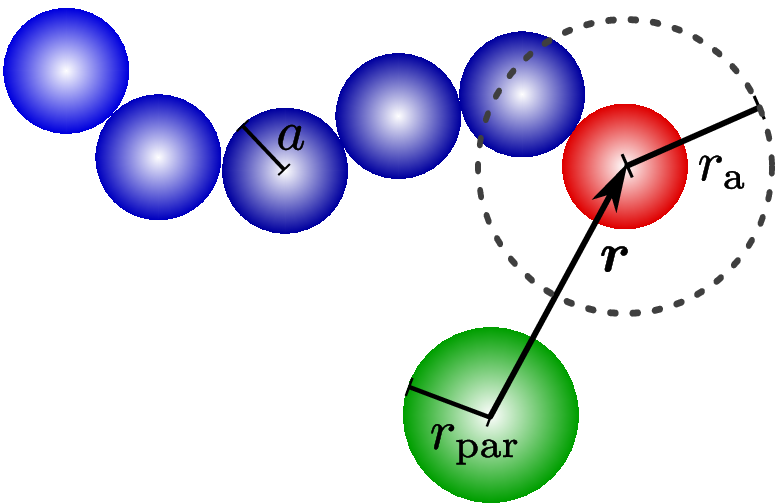}
      \caption{	Schematic view of a bead-spring polymer with a reactive
        end and a free particle of radius $r_\tx{par}$.  The
        particle is absorbed as soon as the separation $r$
        reaches the absorption radius $r_\tx{a}=r_\tx{par}+a$.}
      \label{Fig:polymer-particle}
   \end{center}
\end{figure}

The diffusion-limited reaction we consider is schematically illustrated in \reffig{Fig:polymer-particle}: free particles are
absorbed by the reactive end-monomer of a chain as soon as the separation $r$ reaches the absorption radius $r_\tx{a}=r_\tx{par}+a$.
The association rate is calculated from the renewal approach in
Sec.~\ref{Sec:FPTandRates}, with the only input being the radial
Green's functions for the relative motion of the particle and the chain
end (Eqs.~\ref{Eq:GFrad} and \ref{Eq:VarGFhPolPar}).

\begin{figure}
   \begin{center}
     \includegraphics*[width=3.375in]{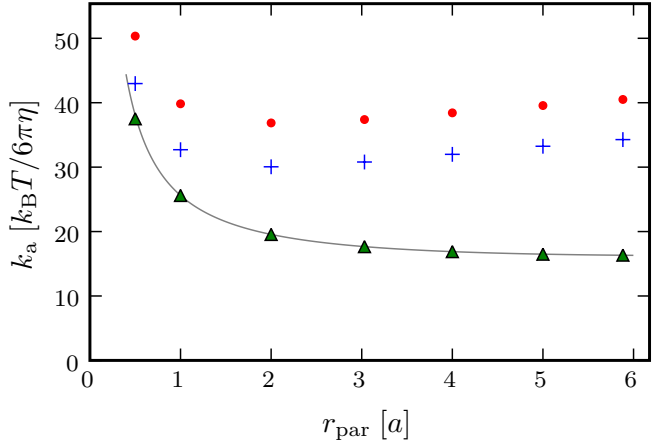}
      \caption[toc entry]{Three estimates for the association rate of
        a free particle to the end-point of a polymer with $L=1000a$,
        $l_\tx{p} = 50a$, as a function of particle radius $r_\tx{par}$.
        For DNA, where $a \approx 1$ nm and $l_\tx{p} \approx 50$ nm, the
        corresponding strand would have a length $L = 1$ $\mu$m.  The rates are
        measured in units of $\kBT/6\pi\eta$, or $\approx 1.3\times
        10^8$ M$^{-1}$s$^{-1}$ for water at room temperature.\textit{ Green triangles} and \textit{solid line} are the Smoluchowski rate considering only the center-of-mass diffusion of the particle and the polymer coil, without
        hydrodynamic coupling (\textit{Green triangles} are based on \reftwoeqs{Eq:GFrad}{Eq:VarGFn2Par}, the \textit{line} denotes \refeq{Eq:Smoluchowski-RatePolPar}). \textit{Red dots}: the rates including the
        internal fluctuations of the polymer, without hydrodynamics (\reftwoeqs{Eq:GFrad}{Eq:VarGFnPolPar}).
        \textit{Blue crosses}: the total rates including both fluctuations and
        hydrodynamics (\reftwoeqs{Eq:GFrad}{Eq:VarGFhPolPar}). All the data marked by \textit{symbols} is obtained
        from the numerical evaluation of the renewal approach (Eqs.~\ref{Eq:RenewalLT}-\ref{Eq:RenewalAsymptoticRates}).}
      \label{Fig:rate-estimates}
   \end{center}
\end{figure}

Since polymer fluctuations and the polymer-particle hydrodynamic
coupling have competing effects on the association rate, it will be
instructive to start with the simple case where the hydrodynamic coupling has been
turned off, i.e.~we use the non-hydrodynamic Green's function
$G_\tx{rad}^\tx{n}$ of \reftwoeqs{Eq:GFrad}{Eq:VarGFnPolPar}.  The resulting association rate as a function
of particle radius $r_\tx{par}$ for a polymer with $L=1000a$, $l_\tx{p} =
50a$ is marked by red dots in \reffig{Fig:rate-estimates}.
  (For DNA,
where $a \approx 1$ nm and $l_\tx{p} \approx 50$ nm, this would correspond
to a strand of length 1 $\mu$m.)  The Smoluchowski rate,
\begin{equation}
\label{Eq:Smoluchowski-RatePolPar}
k_\tx{S}=4\pi(D_\tx{pol}+D_\tx{par})r_\tx{a},
\end{equation}
which involves only the center-of-mass motion of the polymer coil and particle, is shown as a line with the polymer diffusion constant $D_\tx{pol}$ taken from the MFT expansion in \ref{Sec:MFT}. The corresponding numerical results of the renewal approach shown as green triangles are based on \reftwoeqs{Eq:GFrad}{Eq:VarGFn2Par} and again using $D_\tx{pol}$ from \ref{Sec:MFT}. The Smoluchowski value is the standard point of reference when considering DNA-protein interaction~\cite{vonHippel1989}. Clearly, internal polymer fluctuations have a significant impact, enhancing the association rates by $50 - 135\%$ relative to $k_\tx{S}$ over the range $r_\tx{par} = 0.5a - 6a$; the minimum at $r_\tx{par}\approx 2a$ for the rates with internal polymer motion indicates that the effect of larger absorption radius $r_\tx{a}=r_{\text{par}}+a$ with increasing $r_\tx{par}$ quickly dominates the one of decreasing diffusion constant $D_{\text{par}}\propto r_{\text{par}}^{-1}$.

In the absence of hydrodynamics, we can use an earlier result of Berg
\cite{Berg1984} to do a consistency check on the renewal approach
derivation of the association rates.  For the case of a Gaussian
Green's function, and assuming a Gaussian sink profile
$S(r)=\exp{\left(-3r^2/2r_\tx{a}^2\right)}$ instead of a perfectly
absorbing boundary at $r_\tx{a}$, Berg derived an expression for the
association rate in terms of the relative non-hydrodynamic variance
$V^\tx{n}(t)$:
\begin{equation}
\label{Eq:RateBerg}
 k_\tx{a}^\tx{Berg}=(2\pi)^{3/2}\left[\int_0^\infty dt\,\left(V^\tx{n}(t)+2/3\,r_\tx{a}^2\right)^{-3/2}\right]^{-1}.
\end{equation}
Plugging $V^\tx{n}(t)$ from \refeq{Eq:VarGFnPolPar} into
\refeq{Eq:RateBerg}, we recover the non-hydrodynamic rates shown in
\reffig{Fig:rate-estimates} within a difference of 5\% (comparison not shown).

When hydrodynamic interactions are included together with internal
polymer motion ($G^\tx{h}_\tx{rad}$ of \reftwoeqs{Eq:GFrad}{Eq:VarGFhPolPar} is used instead of
$G^\tx{n}_\tx{rad}$ of \reftwoeqs{Eq:GFrad}{Eq:VarGFnPolPar}), the association rates are smaller than in the
non-hydrodynamic case, as seen in the data marked by blue crosses in
\reffig{Fig:rate-estimates}.  This is due to the inhibited mobility
between the particle and the target at short distances.  However the rate
decrease is only $\sim 15-20\%$, so the overall association rate is
still $30-100\%$ larger than the Smoluchowski result for the range of
particle sizes considered.  The magnitude of the rate decrease is
comparable to previous estimates derived for the simpler problem of a
reaction between two spherical particles: Friedman obtained a $15\%$
reduction due to hydrodynamics~\cite{Friedman1966}, Deutch and
Felderhof saw a $46\%$ drop-off~\cite{Deutch1973}, and Wolynes and Deutch predicted a decrease by $29\%$~\cite{Wolynes1976}.

\begin{figure}
   \begin{center}
      \includegraphics*[width=3.375in]{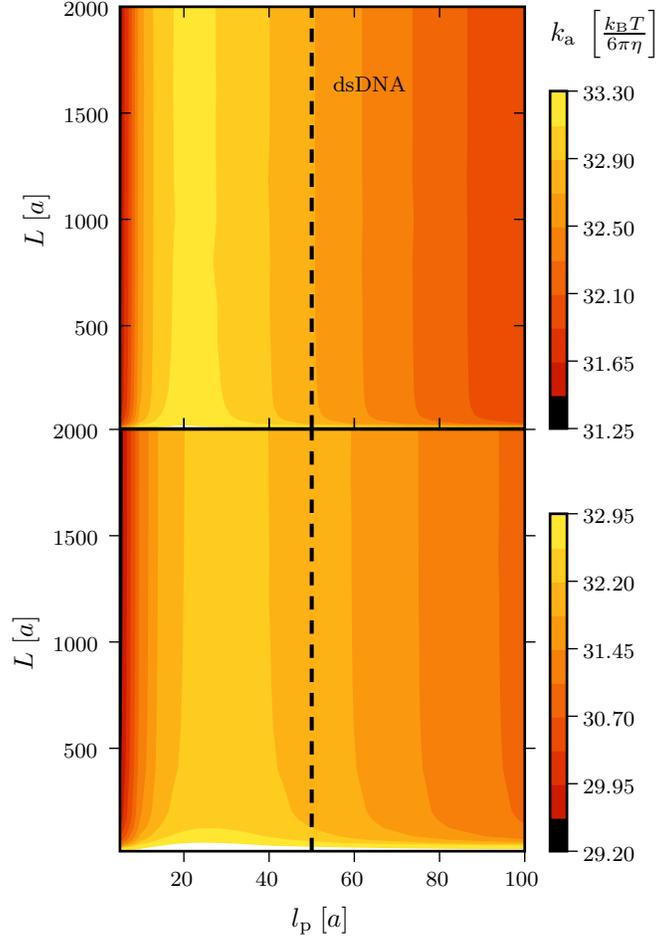}\\[1em]
      \caption{The association rate $k_\tx{a}$ (Eqs.~\ref{Eq:RenewalLT}-\ref{Eq:RenewalAsymptoticRates} with \reftwoeqs{Eq:GFrad}{Eq:VarGFhPolPar}) between a free particle and a polymer end as it varies with the polymer contour length $L$
        and persistence length $l_\tx{p}$ for particle radius
        $r_\tx{par} = a$ (\textit{top panel}) and for $r_\tx{par} = 4a$ (\textit{bottom panel}).
        The rates include the effects of polymer fluctuations and
        hydrodynamic coupling, and are measured in units of
        $\kBT/6\pi\eta$, or $\approx 1.3\times10^8$ M$^{-1}$s$^{-1}$
        in water at room temperature.}
      \label{Fig:contour_abs}
   \end{center}
\end{figure}

\begin{figure}
   \begin{center}
      \includegraphics*[width=3.375in]{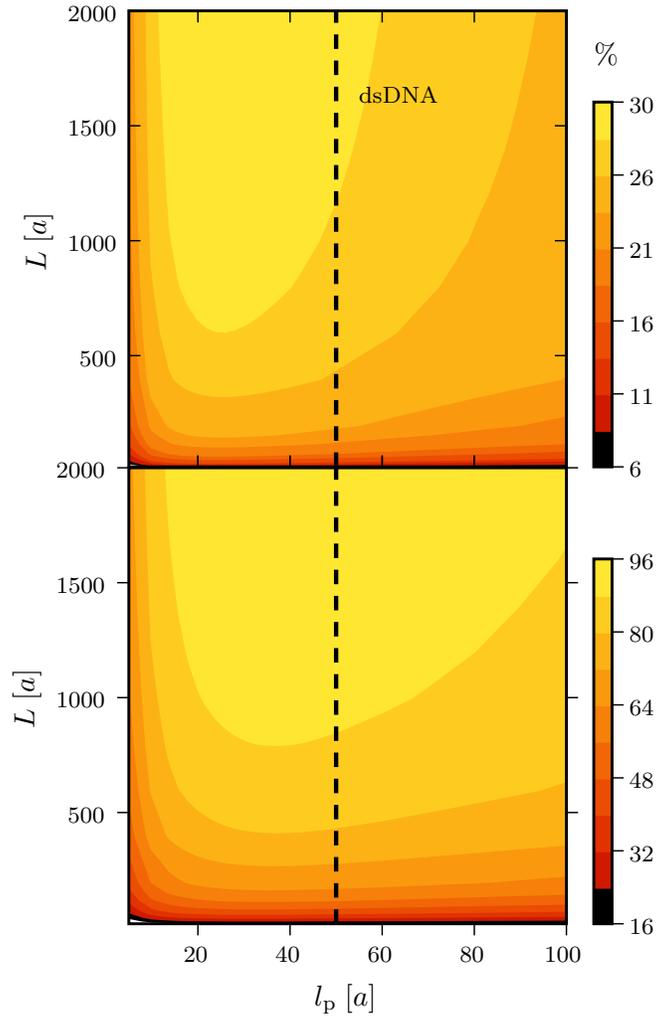}\\[1em]
      \caption{Same as in \reffig{Fig:contour_abs}, but showing the
        percent rate increase over the Smoluchowski result,
        $k_\tx{S}$ (\refeq{Eq:Smoluchowski-RatePolPar}), instead
        of the absolute rate.}
      \label{Fig:contour_rel}
   \end{center}
\end{figure}

Since the conformational fluctuations depend on the mechanical properties of
the chain, it is natural that the association rates will vary with the
parameters $L$ and $l_\tx{p}$ that characterize the semiflexible polymer.
In \reffig{Fig:contour_abs} we show contour diagrams of the
association rate in terms of $L/a$ and $l_\tx{p}/a$ for two different
particle radii: $r_\tx{par}=a$ in the top panel, as would be the
case in a polymerization process, and $r_\tx{par}=4a$ in the bottom
panel, which corresponds to an average protein size for the case of
DNA-protein interaction (with $a=1\text{nm}$). \reffig{Fig:contour_rel}
depicts the same data, but in terms of percent rate increase over the
Smoluchowski result $k_\tx{S}$ of \refeq{Eq:Smoluchowski-RatePolPar}.
 Interestingly, while the absolute rate in
\reffig{Fig:contour_abs} is almost independent of $L$ for $L \gg l_\tx{p}$,
it does show a slight maximum for $l_\tx{p} \approx 20a$ when $r_\tx{par}
=a$ and $l_\tx{p} \approx 30a$ when $r_\tx{par} =4a$. This behavior as a function of $l_\tx{p}$ is an intrinsic property of the polymer, because it
appears also in the absence of hydrodynamic coupling (data not shown). This means that there is an optimal persistence length $l_\tx{p}$ at which the chain reactivity is maximised, and interestingly this optimum is not far from the actual DNA persistence length of $l_\tx{p}=50~\tx{nm}$. The maximum is
more prominent in the relative rates of \reffig{Fig:contour_rel},
where the increase compared to $k_\tx{S}$ rises sharply to $\sim 30\%$
($r_\tx{par} = a$) and $\sim 100\%$ ($r_\tx{par}=4a$) for long
polymers with $l_\tx{p} = 20a - 60a$ including double-stranded DNA ($l_\tx{p}\approx 50a$).

\section{Discussion and summary}
\label{Sec:Discussion}

In this paper, we have studied the association of free particles and a
target site on a semiflexible polymer---a class of diffusion-limited
reactions of wide biological relevance, whether in polymerization of
biomolecules or gene regulation by DNA-binding proteins.  We focused
on two competing effects that are, with the lone exception of the segmental diffusion considered in Ref.~\cite{Berg1984}, entirely neglected in existing
theories for these processes: the bending fluctuations of the polymer
in equilibrium, which enhance the association rate, and the
hydrodynamics between the polymer and particle, which reduces
it.  Quantifying the fluctuations required an accurate description of
internal polymer motion, available through the mean field theory of
semiflexible polymer dynamics.  For the hydrodynamics, we developed a
simple heuristic estimate to model the decrease in mobility when two
diffusing objects approach each other.  The end result of the
competition sensitively depends on the mechanical properties of the
polymer and the size of the reactive particle: we see a maximal
increase over the Smoluchowski rate for persistence lengths $l_\tx{p} = 20a
- 60a$; this increase is $\sim 30\%$ for the case of small particles,
and $\sim 100\%$ for particle sizes typical of regulatory proteins.
As a function of particle size, the reaction rate displays a minimum at a particle radius equal to the polymer diameter; for larger proteins the probability of hitting the DNA increases.

Although based on experimental evidence it is generally assumed that
the binding of proteins to DNA is a diffusion-limited
process~\cite{Halford2004} (and Refs. therein), recent studies in the
slightly different context of protein assisted interior loop formation
of DNA~\cite{Polikanov2007,Catto2008} indicate that this assumption is
probably not justified for all kinds of DNA-protein systems. Within
our approach we have neglected the possibility of encounters between
the protein and the target site on the DNA that do not immediately
lead to a reaction, for example due to orientational constraints. This
is clearly an oversimplification, which however can be approximately
rectified by a reaction factor $\Gamma$ (see \ref{Sec:AddEffects}).

\begin{figure}
   \begin{center}
      \includegraphics*[width=3.375in]{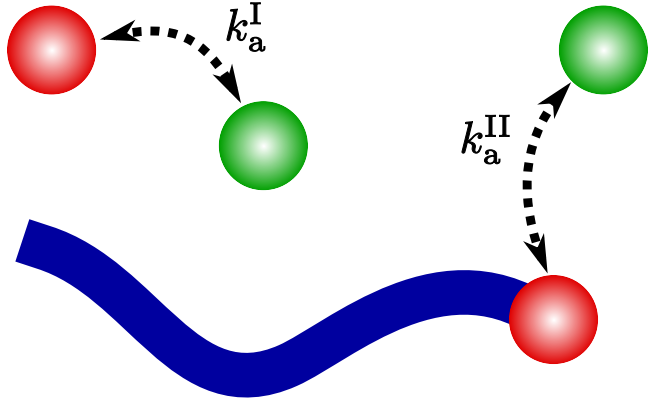}
       \caption{Schematic representation of two scenarios for a
         bimolecular reaction: I) both species are free in solution;
         II) one of the species is attached to the end of a semiflexible
         polymer.  The comparison of the diffusion-limited association
         rates $k_\tx{a}^\tx{I}$ and $k_\tx{a}^\tx{II}$ should highlight the role of
         internal polymer motion in such reactions.}
      \label{Fig:exp_rates}
   \end{center}
\end{figure}

Directly testing our theoretical predictions may be possible with
existing experimental techniques.  In particular, one can study the
role of polymer fluctuations on association rates by comparing two
different setups involving bimolecular diffusion-limited reactions,
shown schematically in \reffig{Fig:exp_rates}.
  In scenario I both
reactants are free in solution, while in scenario II one of the
reactants is bound to the end of a semiflexible polymer. The
corresponding rates $k_\tx{a}^\tx{I}$ and $k_\tx{a}^\tx{II}$ could be measured through
simple kinetic experiments or using fluorescence microscopy.  In the
latter case, the reactants would be fluorophores and quencher
molecules, and the rate of the diffusion-limited quenching reaction
can be extracted from the decay of the total fluorescence signal over
time.  As shown in Table \ref{Tab:exp_rates}, our theory predicts a
ratio $k_\tx{a}^\tx{II}/k_\tx{a}^\tx{I} \approx 0.77$ for the two scenarios using
$r_\tx{par} =a$ and a 1 $\mu$m strand of DNA.
\begin{table}
   \begin{center}
	\begin{tabular}[t]{l|ccc}
		Rate estimates including & $k_\tx{a}^\tx{I}$ &  $k_\tx{a}^\tx{II}$ & $k_\tx{a}^\tx{II}/k_\tx{a}^\tx{I}$\\
		\hline
		center of mass motion & 50.27 & 25.14 & 0.5\\
		plus internal polymer motion & - & 39.83 & -\\
 		plus hydrodynamics & 42.52 & 32.71 & 0.77
	\end{tabular}
	\caption{Estimates for diffusion-controlled bimolecular
          association rates in the experimental scenarios shown in
          \reffig{Fig:exp_rates}.  The polymer is assumed to be a 1
          $\mu$m strand of dsDNA, and the particle radius
          $r_\tx{par} = 1$ nm.  The rates $k_\tx{a}$ are given in units
          of $\kBT/6\pi\eta$, or $\approx 1.3\times10^8$
          M$^{-1}$s$^{-1}$ in water at room temperature.}
      \label{Tab:exp_rates}
   \end{center}
\end{table}
This is clearly distinguishable from the Smoluchowski prediction, $k_\tx{a}^\tx{II}/k_\tx{a}^\tx{I} =
0.5$, which ignores hydrodynamics and internal fluctuations.

\section{Acknowledgements}
The authors would like to thank Ana-Maria Florescu and Marc Joyeux for useful discussions, and the DFG (grants NE 810/7 and SFB 863) for financial support.

\appendix
\renewcommand{\thesection}{Appendix\,\Alph{section}}
\renewcommand{\theequation}{\Alph{section}\arabic{equation}}
\setcounter{equation}{0}

\setcounter{equation}{0}
\section{Additional Effects on the Association Rates}
\label{Sec:AddEffects}

Throughout this paper we have considered an idealized picture of the
association process assuming that a reaction takes place as soon as
the radial distance $r$ reaches the absorption radius $r_\tx{a}$. A
more realistic description should take into account a number of other
factors: (i) molecules are in general not equally reactive on their
entire surface, i.e. binding also depends on the relative orientation
of the reactants; (ii) the correct relative orientation alone might
not directly lead to binding due to additional requirements,
e.g. conformational changes of the molecules. In addition we have
assumed that the polymer-particle dynamics is purely diffusive;
deviations from this behavior might be caused by (iii) excluded-volume
effects between the particle and non-target monomers of the polymer
together with (iv) electrostatic interactions between the particle and
polymer. Due to all these effects, the real rate
$k_\tx{a}^\tx{real}=\Gamma k_\tx{a}$ will differ from our predicted
rate $k_\tx{a}$ by a factor $\Gamma$. Below we give a rough estimate
of $\Gamma$ for each of these cases and argue that, though modifying
the absolute values of the association rates, these considerations do
not alter the main qualitative conclusions of our work.

(i) Rate-influencing effects of orientational constraints have been
examined for the case of diffusion-controlled bimolecular reactions of
spherical particles~\cite{Solc1971, Shoup1981}; the general ideas can
also be applied to our polymer-particle system. If only a fraction $p$
of the target's surface is reactive, the most important result is that
the association rate is \textit{not} simply reduced by $p$. Rather the
factor $\Gamma>p$ reflects the fact that a first encounter between
the reactants involving inert parts of the target's surface (occuring
with probability $1-p$) does not exclude the possibility of a
successful reaction at a later time.  In the limit where the process
of mutual reorientation is fast compared to the time-scale of relative
diffusion, one even recovers $\Gamma=1$: the partial surface
reactivity is then completely compensated for by fast orientational
changes of the target.  End-tangent fluctuations of semiflexible
polymers are typically fast---the relaxation times of the high
frequency normal modes are on the order of $\sim a^2/(\mu_0 \kBT)
\approx 4~\text{ns}$.  Plugging this timescale into the method of
Ref.~\cite{Shoup1981}, one can extract representative values for
$\Gamma$: when $p=0.5$ we find $\Gamma\approx 0.79$ for $r_\tx{par}=a$
and $\Gamma \approx 0.91$ for $r_\tx{par}=5a$.

(ii) For the case where the reaction is limited by the conformational
state of the particle, the timescale of conformational transitions is
typically much larger than the timescale of relative diffusion between
the particle and the target.  Where such a separation of timescales
exists, the rate reduction factor $\Gamma$ is approximately just the
probability for the particle to be in the correct conformation.

(iii) Throughout this paper excluded-volume effects between polymer
and particle have been neglected. The diffusion towards the target is
altered by the presence of the polymer coil; however, since the
effective segment density is low for a semiflexible polymer in good
solvent, we only expect minor corrections. On the other hand, as can
be seen from \reffig{Fig:polymer-particle}, neighboring monomers to
the target monomer exclude a solid angle $\Omega^\tx{ex}$ from the
target's surface. In the case of the end-monomer being the target, the
excluded solid angle is simply
\begin{equation}
\Omega^\tx{ex}=\int_{0}^{2\pi}\tx{d}\phi\,\int_{0}^{\theta^\tx{max}}\tx{d}\theta\,\sin\theta=2\pi\frac{r_\tx{par}}{r_\tx{par}+a},
\end{equation}
with $\theta^\tx{max}=\arccos(r_\tx{par}/(r_\tx{par}+a))$. The surface
of the target-monomer over which a reaction can take place is
therefore reduced by a factor $p=1-\frac{\Omega^\tx{ex}}{4\pi}$
ranging from $1$ in the limit $r_\tx{par}\to0$ to $1/2$ in the limit
$r_\tx{par}\to\infty$. At first glance the situation may appear
similar to case (i), where a fraction $1-p$ of the
target surface is inert.  However here the $1-p$ fraction is
inaccessible due to excluded volume; since the particle cannot get
near to that portion of the target, the rate reduction will be less
drastic (for a given $p$, the corresponding $\Gamma$ value will be
closer to 1).  

(iv) For electrostatic effects, which play a role, for example, in the
non-specific interactions between proteins and DNA, the value of
$\Gamma$ is difficult to estimate analytically.  An effective method
for determining $\Gamma$ using BD simulations has been proposed in
Ref.~\cite{Northrup1984}; the authors of that study considered
attractive electrostatics between oppositely charged monovalent ions
in water.  For an unscreened Coulomb potential $\Gamma=6.82$ and
$7.31$, respectively, with and without hydrodynamic
interactions on the level of the Oseen tensor.  The corresponding
values for a Debye length of 1 nm were $\Gamma = 4.40$ and $4.80$.

The total value of $\Gamma$ seen in an experimental situation is
expected to reflect some combination of the above effects.  While this
will modify the absolute association rates, the relative rate changes
discussed in our work should remain valid: in all scenarios, when we
compare to a fixed target in the free-draining limit, the inclusion of
hydrodynamics should retard the reaction, but should be more than compensated for
by the speed-up due to internal polymer fluctuations.

\section{Mean field theory for semiflexible polymer dynamics}
\label{Sec:MFT}
The simplest description of a semiflexible polymer is the worm-like
chain model: the polymer is represented by a continuous,
differentiable space curve $\V r(s)$ of contour length $L$, where the
arc-length variable $s$ ranges from $-L/2$ to $L/2$.  The associated
elastic energy $U_\tx{WLC}$, the continuum analogue of \refeq{Eq:PotentialsBD}, is given by~\cite{Kratky1949}:
\begin{equation}
\label{Eq:BendingWLC}
 U_\tx{WLC}[\V r(s)]=\frac{\kappa}{2}\int_{-L/2}^{L/2} ds\,\left(\pdiffarg{\V
   u(s)}{s}\right)^2.
\end{equation}
The bending rigidity is $\kappa=l_\tx{p}\kBT$, and the tangent vector $\V
u=\partial \V r/\partial s$ is constrained by local inextensibility to
unit length, $\V u^2(s)=1$ at each $s$.  Since this constraint leads
to nonlinear equations of motion, an alternative, approximate model is
required.  Within the mean field theory approach~\cite{Winkler1994,
  Ha1995} the local constraint is relaxed and replaced by the global
and end-point conditions $\Av{\int ds\,\V u^2(s)}=L$ and $\Av{\V
  u^2(\pm L/2)}=1$ . The result is a Gaussian mean field Hamiltonian
which incorporates a finite extensibility in addition to the bending
term:
\begin{equation}
\label{Eq:HamiltonianMFT}
\begin{split}
 U_\tx{MF}[\V r(s)]=\frac{\epsilon}{2}\int_{-L/2}^{L/2} ds\,\left(\pdiffarg{\V u(s)}{s}\right)^2+\nu\int_{-L/2}^{L/2} ds\,\V u^2(s)+\nu_0(\V u^2(L/2)+\V u^2(-L/2)),
\end{split}
\end{equation}
where $\epsilon=3l_\tx{p} \kBT/ 2$, $\nu=3\kBT/(4l_\tx{p})$, and
$\nu_0=3\kBT/4$.  In this form the Gaussian model exactly
reproduces various lowest-order equilibrium averages of the worm-like chain,
most importantly the tangent-tangent correlation function, and other derived quantities such as the mean
square end-to-end distance.

The dynamic theory for the Gaussian semiflexible polymer is based on
the hydrodynamic pre-averaging approach of Ref.~\cite{Harnau1996},
analogous to that used for the Zimm model~\cite{Zimm1956} in the case
of flexible chains.  The time evolution of a point $s$ on the polymer
contour is governed by the Langevin equation:
\begin{equation}
\label{Eq:LangevinMFT}
\begin{split}
\pdiff{s}\V r(s,t)=-\int_{-L/2}^{L/2}ds'\,\T\mu_\tx{avg}(s,s')\,\frac{\delta U_\tx{MF}}{\delta \V r(s',t)}+\V
\xi(s,t),\\ \Av{\V\xi(s,t)\otimes\V\xi(s',t')}=2\kBT\,\T\mu_\tx{avg}(s,s')\,\delta(t-t').
\end{split}
\end{equation}
Here we use the pre-averaged mobility tensor $\T
\mu_\tx{avg}(s,s^\prime)$, which is obtained from the standard
Rotne-Prager tensor by averaging over all equilibrium configurations
of the polymer.  As seen in \refeq{Eq:RotnePragerTensor}, the
original Rotne-Prager mobility involves a dependence on the spatial
distance between polymer points, and hence on the specific
configuration of the chain.  This would lead to nonlinear equations
of motion, a problem which is resolved in the pre-averaging
approximation, where the mobility depends only on the arc-length
coordinates $s$ and $s^\prime$.  For the mean field model of
\refeq{Eq:HamiltonianMFT} the pre-averaged mobility has the
form~\cite{Harnau1996}:
\begin{equation}\label{Eq:muavg}
\begin{split}
\T\mu_\tx{avg}(s,s^\prime) = &\Biggl[ 2a\mu_0 \delta(s-s^\prime) +\frac{\Theta(|s-s^\prime|-2a)}{\eta\sqrt{6\pi^3 \sigma(|s-s^\prime|)}} \exp\left(-\frac{6a^2}{\sigma(|s-s^\prime|)}\right)\Biggr]\unitmatrix,
\end{split}
\end{equation}
where $\sigma(l) = 2l_\tx{p} l - 2l_\tx{p}^2 (1-\exp(-l/l_\tx{p}))$.  The microscopic
length scale $a$ in the continuum theory corresponds to the monomer
radius in the discrete BD simulations, and the unit step
function $\Theta$ in \refeq{Eq:muavg} serves as a short-distance
cutoff for the hydrodynamic interactions.

The pre-averaged Langevin equation can be solved through a normal mode
decomposition, with the eigenmodes fulfilling free-end boundary
conditions at $s=\pm L/2$.  This reduces \refeq{Eq:LangevinMFT} to a
set of ordinary differential equations coupled by a hydrodynamic
interaction matrix; the diagonalization of this matrix yields simple
Langevin equations for the decoupled normal mode amplitudes $\V
P_n(t)$ (with stochastic contributions $\V Q_n(t)$):
\begin{equation}
\label{Eq:LangevinPn}
\begin{split}
&\pdiff{t} \V P_0(t)=\mathbf Q_0(t), \qquad \pdiff{t} \V P_n(t)=-\Lambda_n\V P_n(t)+\mathbf Q_n(t),\quad n=1,\dots,N,\\
&\langle Q_{ni}(t)Q_{mj}(t')\rangle=\,2\kBT\delta_{ij}\delta(t-t')\Theta_n\delta_{nm}.
\end{split}
\end{equation}
The vectors $\V P_n(t)$ and $\V Q_n(t)$ are related to $\V r(s,t)$ and the
stochastic velocities $\V \xi(s,t)$ through the expansions $\V r(s,t)
= \sum_{n=0}^N \V P_n(t) \Psi_n(s)$ and $\V \xi(s,t) = \sum_{n=0}^N \V
Q_n(t) \Psi_n(s)$, where the scalar functions $\Psi_n(s)$ are the decoupled normal modes.
The modes are ordered in such a way that the eigenvalues $\Lambda_n$
(inverse relaxation times) increase with $n$.  Following
Ref.~\cite{Hinczewski2009}, we set the high-frequency cutoff $N$ for
the mode number to $N = \lceil L/8a \rceil$, which was shown to
give good agreement at short times with BD simulations.  At longer
times, where the polymer fluctuations are at length scales much larger
than the monomer radius $a$, the dynamics does not depend on the
precise choice of the cutoff.  $\Lambda_n$ and the fluctuation-dissipation
parameters $\Theta_n$ can be directly derived from the tensor
$\T\mu_\tx{avg}$ evaluated numerically in the normal mode basis.
Full details of this procedure, together with the explicit form of the
normal modes $\Psi_n(s)$, are given in Ref.~\cite{Hinczewski2009}.

\refeq{Eq:LangevinPn} for a given mode $n$ can be mapped onto the
well-known problem of a particle diffusing with friction constant
$\zeta$ in a harmonic potential of strength $k$ centered at $\V x=0$
\cite{DoiEdwards}.  We identify the mode amplitude $\V P_n(t)$ with the
position $\V x$ of the particle and set $\Lambda_n=k/\zeta$,
$\Theta_n=1/\zeta$.  This allows us to calculate the explicit form of
the Green's function for the diffusive process, in other words the
transition probability $P(\V P_n(t),\V P_n(0);t)$ for observing the
amplitude $\V P_n(t)$ after time $t$, starting from the initial
amplitude $\V P_n(0)$:
\begin{equation}
\label{Eq:GFNormalModes}
P(\V P_n(t), \V P_n(0);t)=\left(2\pi \sigma^2_n(t)\right)^{-3/2}\,
\exp\left(-\frac{(\V P_n(t)-\V M_n(t))^2}{2 \sigma^2_n(t)}\right).
\end{equation}
The probability has the form of a spreading Gaussian, with
time-dependent mean $\V M_n(t)$ and variance $\sigma_n^2(t)$ given by:
\begin{align}
\label{Eq:MeanGFNormalModes}
\V M_0(t)&=\V P_0(0), \qquad \V M_n(t)=\V P_n(0)e^{-\Lambda_n t},\quad\forall n>0, \\
\label{Eq:VarGFNormalModes}
\sigma^2_0(t)&=2 L D_\tx{pol} t, \qquad \sigma^2_n(t)=\kBT \Theta_n\Lambda_n^{-1}(1-e^{-2\Lambda_n t}),\quad \forall n>0.
\end{align}
Here
\begin{equation}
D_\tx{pol}\equiv\kBT\Theta_0\Psi_0(s)^2=\kBT\Theta_0\frac{1}{L},
\end{equation}
denotes the center-of-mass diffusion constant of the polymer coil.  Modes with different mode
number $n$ evolve independently of each other.  Transition
probabilities between different polymer configurations, each of them
corresponding to a certain (unique) set of normal mode amplitudes
$\{\V P_n(t)\}$, are therefore expressed as the product of the
transition probabilities for the individual modes. As we are
interested in the motion of a single point $s$ on the polymer contour,
the integrations over initial configurations with $\V r(s,0)=\V r_0$
and final configurations with $\V r(s,t)=\V r$ can be readily
performed yielding the Green's function in Eqs.~\ref{Eq:GF3D} and \ref{Eq:VarGFPol}.

\setcounter{equation}{0}
\section{Hydrodynamic interactions}
\label{Sec:Hydro}
In \ref{Sec:MFT} we were able to describe the long-range
hydrodynamic interactions between various points on a polymer coil,
and their influence on the internal relaxation of the chain.  A key
simplifying feature in this analysis is the fact that the spatial
relationship between any two points on the contour is constrained:
their hydrodynamic interactions as the polymer fluctuates in equilibrium can be
taken into account through the pre-averaging approximation.  For the
case of a free particle and a target site on the polymer coil,
estimating hydrodynamic interactions is more difficult: the particle
can drift away, and the strength of the interaction will be highly
dependent on the initial conditions and the elapsed time.  To
understand the role of hydrodynamics in this situation, we first
consider the simpler case of two freely diffusing spherical particles,
before tackling the full problem of polymer-particle hydrodynamics.
Though it may seem trivial, even the two particle case presents a challenging problem and can only be
dealt with approximately~\cite{Friedman1966,Deutch1973}.

\subsection*{Case of two spherical particles}
\label{Sec:Hydro2Par}
We consider two non-interacting particles at positions $\V r_i(t)$, $i=1,2$, described by
the Langevin equations:
\begin{equation}
\label{Eq:Langevin2Par}
\begin{split}
\pdiff{t}\V r_i(t)&=\V\xi_i(t),\\
\Av{\V\xi_i(t)\otimes\V\xi_j(t')}&= 2 \kBT \,\delta(t-t')\,\left[\delta_{ij}\mu_i\unitmatrix+(1-\delta_{ij})\T\mu(\V r_{12})\right],
\end{split}
\end{equation}
where $\mu_{i}$ is the self-mobility of particle $i$, and
hydrodynamics are expressed through the Rotne-Prager tensor
$\T\mu(\V r_{12})$, defined in \refeq{Eq:RotnePragerTensor}, dependent on the inter-particle separation $\V
r_{12}$. Hydrodynamic interactions are long-ranged ($\propto
r_{12}^{-1}$) and therefore negligible only at distances much larger
than the sum of the particle radii, $r_{12}\gg a_1+a_2$. For small
separations, the stochastic motion of the particles is highly
correlated, leading to a decrease of their relative mobility.  To
get a realistic estimate of binding rates, where particles clearly
have to approach each other, it is therefore necessary to take these
hydrodynamic effects into consideration.

The main quantity of interest is the radial Green's function for the
relative motion of the particles, the probability that two particles
starting from a distance $r_0=\abs{\V r_{12}(0)}$ reach a distance
$r=\abs{\V r_{12}(t)}$ in time $t$.  For comparison, we will consider
the situation both with and without hydrodynamic interactions
, labeling the respective Green's function $G^\tx{h}_\tx{rad}$ and $G^\tx{n}_\tx{rad}$
. The second case, where the particles are
totally decoupled and only the self-mobilities enter into the
stochastic correlations of \refeq{Eq:Langevin2Par}, is trivial and leads to the functional form of Eq.~\ref{Eq:GFrad} with the variance of Eq.~\ref{Eq:VarGFn2Par}, where $D_i=\mu_i \kBT$ is the diffusion constant of particle $i$.
Note that the variance $V^\tx{n}(t)$ is just $1/3$ of the MSD
$\langle\left(\V r_{12}(t)-\V r_{12}(0)\right)^2\rangle$.

In contrast, for the hydrodynamic case one cannot derive an analytical
form for $G^\tx{h}_\tx{rad}$.  Thus we will have to resort to a
heuristic approximation: we assume $G^\tx{h}_\tx{rad}$ has the
same functional form as \refeq{Eq:GFrad}, but with a different
variance $V^\tx{h}(t)$, reflecting the slower relative motion of the
particles.  Since this variance is related to the MSD of $\V
r_{12}(t)$, we begin by evaluating this MSD.  From
\refeq{Eq:Langevin2Par} and the definition of the Rotne-Prager tensor,
\refeq{Eq:RotnePragerTensor}, one can obtain the following exact
relationship:
\begin{equation}
\label{Eq:MSDDistanceVectorHydro}
\langle\left(\V r_{12}(t)-\V r_{12}(0)\right)^2\rangle=6(D_1+D_2)t-12\underbrace{\kBT\Av{\int_{0}^{t}dt'\,\frac{1}{6\pi\eta r_{12}(t')}} }_{\equiv\chi t}.
\end{equation}
The second term on the right can be rewritten as $-12\chi t$,
defining a time-dependent coupling parameter $\chi$, which quantifies
the slow-down in relative diffusion compared to the non-hydrodynamic
case.  This parameter involves both a time and ensemble average over
the trajectory $\V r_{12}(t)$, and hence is also dependent on the
initial separation $r_0$.  Our heuristic approach approximates $\chi$
by an effective coupling function $\bar\chi(r_0,t)$, where the time
and ensemble averages have been replaced by a single ensemble average
involving the Green's function $G^\tx{n}_\tx{rad}$ for the non-hydrodynamic
system:
\begin{equation}
\label{Eq:ChiBar2Par}
\begin{split}
\bar\chi(r_0,t)=\frac{\kBT}{6\pi\eta}\int_{r_a}^{\infty} dr\,\frac{1}{r}G^\tx{n}_\tx{rad}(r,r_0;t)=
\frac{\kBT}{6\pi\eta}\,\frac{1}{2 r_0}\left[\erf{\left(\frac{r_0-r_a}{\sqrt{2V^\tx{n}(t)}}\right)}+\erf{\left(\frac{r_0+r_a}{\sqrt{2V^\tx{n}(t)}}\right)}\right].
\end{split}
\end{equation}
As in the pre-averaging approximation of the previous section,
hydrodynamic interactions are cut off below the distance $r_a = a_1 +
a_2$ where the particles overlap.  This effective parameter shows the
correct limiting behavior: $\bar\chi(r_0,t)\propto r_0^{-1}$ for short
times, where particles are close to their initial separation, and
$\bar\chi(r_0,t) \propto ((D_1+D_2)t)^{-1/2}$ for long times where the
particles have drifted far away from each other.  In analogy to the
non-hydrodynamic case, we define the variance $V^\tx{h}(t)$ as 1/3
of the MSD, using the effective parameter $\bar\chi(r_0,t)$ instead of
$\chi$.  Thus the final form for our approximate hydrodynamic Green's
function is reflected in Eqs.~\ref{Eq:GFrad} and \ref{Eq:VarGFh2Par}.
The role of the effective coupling parameter $\bar\chi$ is to reduce
the relative mobility of the particles when they are near to each
other and thus subject to strong hydrodynamic interactions; in the
long-time limit the hydrodynamic effects become negligible as the
particles move to large separations.

In principle, the procedure outlined above to estimate $\chi$ can be
iterated to produce higher-order approximations: in deriving
$\bar\chi(r_0,t)$ one can use the hydrodynamic Green's function
$G_\tx{rad}^\tx{h}$ of Eqs.~\ref{Eq:GFrad}  and \ref{Eq:VarGFh2Par} instead of
$G_\tx{rad}^\tx{n}$.  This would lead to a better approximation
for $\chi$, and hence a more accurate Green's function which could be
input into the next level of the approximation.  This iterative
method converges quickly, so for simplicity we restrict ourselves to
the first order results.

\subsection*{Case of a DNA-target site and a free particle }
\label{Sec:HydroPolPar}
This approach for two freely diffusing particles can be generalized to
the problem of a free particle and a polymer.  The non-hydrodynamic
case is again simple, with the relative motion of the particle and a
point $s$ on the chain described by a radial Green's function which
has exactly the same form as \refeq{Eq:GFrad}.  The only difference
is the variance $V^\tx{n}(t)$, which now includes the
contribution of the polymer's internal modes as shown in Eq.~\ref{Eq:VarGFnPolPar}.
Here $D_\tx{par} = \kBT /6\pi \eta r_\tx{par}$ is the diffusion
constant of the free particle, and $r_\tx{par}$ is the particle
radius.

For the hydrodynamic case, we can divide the complicated interactions
between the particle and the chain into three parts: (i) the influence
of the local region of radius $a$ around the target site $s$ on the free particle;
(ii) the influence of the rest of the chain on the particle; (iii) the
back influence of the particle on the entire polymer.  For the
specific problem we consider---association rates to a given target
site---hydrodynamics plays a significant role only in the close
vicinity of the target.  Hence contribution (i) will dominate.  The
back influence in (iii) should be negligible for free particles
comparable in size to the monomers in the chain, $r_\tx{par} \sim
{\cal O}(a)$, since the motion of the polymer is mainly governed by
relaxation of the internal modes.  The relative unimportance of (ii)
is more subtle: one can take it into account in a more elaborate
numerical evaluation of the pre-averaged MFT Langevin equations, but
comparison to the simpler approximation discussed below does not show
significant improvement with respect to BD simulations (which include
all three contributions).  Thus we can construct a simple estimate for
the hydrodynamic Green's function by focusing on contribution (i).
$G^\tx{h}_\tx{rad}$ has the same form as the two-particle case,
\refeq{Eq:GFrad}, but with a variance given by Eq.~\ref{Eq:VarGFhPolPar} with
\begin{equation}
\label{Eq:ChiBarPolPar}
 \bar\chi(r_0,t)=\frac{\kBT}{6\pi\eta}\,\frac{1}{2r_0}\left[\erf{\left(\frac{r_0-r_a}{\sqrt{2V^\tx{n}(t)}}\right)}+\erf{\left(\frac{r_0+r_a}{\sqrt{2V^\tx{n}(t)}}\right)}\right],
 \end{equation}
in analogy to \refeq{Eq:VarGFh2Par}.  Here $V^\tx{n}(t)$ is the
non-hydrodynamic polymer-particle variance of \refeq{Eq:VarGFnPolPar}.
In this way we have accounted for both the internal fluctuations of
the target site and, through $\bar\chi(r_0,t)$, the decrease in relative
mobility for the approaching particle.


\end{document}